\documentclass[12pt]{iopart}

\usepackage{iopams}
\usepackage{graphicx,amssymb}
\usepackage{color}

\newtheorem{proposition}{Proposition}
\newtheorem{theorem}{Theorem}
\newtheorem{corollary}{Corollary}

\newcommand{\be}{\begin{equation}}
\newcommand{\ee}{\end{equation}}
\def\tr{\mathop{\rm tr}\nolimits}

\def\dif{{\rm d}}
\def\ci{\mathop{\textrm{i}}\nolimits}

\def\ci{{\rm i}}

\begin{document}
\title[Dimension of the isometry group]
{Dimension of the isometry group in spacetimes with an invariant frame}

\author{Juan Antonio S\'aez$^1$, Salvador Mengual$^{2}$\footnote{Author to whom any correspondence should be adressed.} and Joan Josep Ferrando$^{2,3}$}

\address{$^1$\ Departament de Matem\`atiques per a l'Economia i l'Empresa,
Universitat de Val\`encia, E-46022 Val\`encia, Spain}

\address{$^2$\ Departament d'Astronomia i Astrof\'{\i}sica, Universitat
de Val\`encia, E-46100 Burjassot, Val\`encia, Spain}

\address{$^3$\ Observatori Astron\`omic, Universitat
de Val\`encia, E-46980 Paterna, Val\`encia, Spain}

\ead{juan.a.saez@uv.es; salvador.mengual@uv.es; joan.ferrando@uv.es}

\begin{abstract}
The necessary and sufficient conditions for a spacetime with  an invariant frame to admit a group of isometries of dimension $r$ are given in terms of the connection tensor $H$ associated with this frame. In Petrov-Bel  types I, II and III, and in other spacetimes where an invariant frame algebraically defined by the curvature tensor exists, the connection tensor $H$ is given in terms of the Weyl and Ricci tensors without an explicit determination of the frame. Thus, an IDEAL (intrinsic, deductive, explicit and algorithmic) characterization of these spacetimes follows. Some examples show that this algorithm can be easily implemented on the {\em xAct Mathematica} suite of packages. 
\end{abstract}
%

%


\section{Introduction}

In Relativity, the existence of local isometries (Killing vectors) is one of the most used facts to get a tractable system when dealing with Einstein  equations. The group of isometries enables us to adapt coordinates that can considerably simplify the integration of the field equations. The problem of finding all the Killing vectors admitted by a spacetime or getting a set of invariants characterizing the number of such symmetries  and some properties of their action (transitivity, isotropy group,...) is an old well-established problem in an $n$-dimensional Riemannian space (see for example the classic text by Eisenhart \cite{eisenhart-33}, and reference \cite{kramer} for a Relativistic context).

A Killing vector $\xi$  is (locally) completely determined by its value  $\xi$ at a point ($n$ parameters) and that of its covariant derivative $\nabla \xi$ at the same point ($n (n\!-\!1)/2$ parameters). Then, the maximum freedom in choosing a Killing field is the sum of both, namely $n(n+1)/2$ parameters, which is the maximum number of independent Killing fields admitted by an $n$-dimensional manifold. Constraints on the possible values of the Killing at a point, $\xi$, restrict the dimension of the orbits, and constraints on the values of $\nabla \xi$ restrict the dimension of the isotropy group. 

The equivalence method developed by Cartan \cite{cartan} and adapted to a Lorentzian four-dimensional manifold by Brans \cite{brans} and Karlhede \cite{karlhede} (see also \cite{karlhede-maccallum}) is based on the obtention of scalar invariants associated with the Riemann tensor and its successive derivatives. In the literature, a lot of papers are devoted to discuss the number of invariants necessary to fully define a metric and their construction. For a discussion on the different results one can see the review by MacCallum \cite{MacCallum-2015}, where an exhaustive bibliography on this subject is detailed. This method can also be used to obtain the dimension of the orbits and the number of  Killing fields admitted by a given spacetime \cite{kramer, karlhede-maccallum, tomoda}. Although it is theoretically clearly established, the number of scalars that have to be computed is formidable (see \cite{tomoda} and references therein for a recent review on this subject), and a general algorithm implementing this approach is not known yet. 

Our aim here is to perform a tensorial approach allowing us, when an invariant frame can be determined, to get the dimension of the group of isometries in an algorithmic manner by taking the successive derivatives of the so called
{\it connection tensor} up to the fourth order. A similar approach has been performed in obtaining the dimension of the isometry group in three-dimensional Riemannian spaces \cite{FS-K3}, and in studying the homogeneous three-dimensional  spaces, both Riemannian \cite{FS-G3} and Lorentzian \cite{FS-L3}. Essentially, our algorithm computes the Cartan invariants in the invariant frame without the explicit determination of the frame itself in most of the cases.

In the framework of the Cartan approach, Eisenhart \cite{eisenhart-33} proved that the dimension of the isometry group is $r = n(n+1)/2- m$, where $m$ is the rank of the linear system for $\xi$ and $\nabla \xi$ defined by the integrability conditions of the Killing equations, namely, ${\cal L}_\xi \nabla^p R=0$, $p \geq 0$.

Any Killing field must be orthogonal to the gradient of
every invariant scalar. Consequently, the number $d$ of functionally independent
scalars constrains the dimension of the orbits, $s \leq n - d$. Kerr proved that equality holds for positive definite Riemannian spaces \cite{Kerr-62} and for four-dimensional Einstein spaces \cite{Kerr-63}. 

Here, we consider another case in which the equality also holds: the $n$-dimensional Riemannian spaces where an invariant Riemann-frame exists. An invariant frame $\{e_{a}\}$ that can be built from the Riemann tensor $R$ and its covariant derivatives will be called a {\em Riemann-frame ($R$-frame)}. Since the components of the Riemann tensor in the frame $\{e_{a}\}$ depend on the connection coefficients $\gamma_{ab}^c$ and their first-order directional derivatives $e_d(\gamma_{ab}^c)$, all the Riemann scalar invariants depend on $\gamma_{ab}^c$ and all their successive directional derivatives.

In section \ref{sec-Riemann-frame} we study the spaces where a $R$-frame exists and we show an essential property: if the directional derivatives of the connection coefficients of order $q$ depend on those of lower order, then those of order greater than $q$ also depend on those of lower order. This fact enables us to perform a finite method to determine the dimension of the admitted group of isometries. We finish this section by specifying this method for a four-dimensional spacetime.   

Although finite, the above quoted method involves a colossal number of conditions that makes the elaboration of the algorithm cumbersome. In section \ref{sec-dimension} we solve this shortcoming by introducing, associated with a $R$-frame, the connection tensor and its differential concomitants. They collect all the invariant scalars of the Riemann tensor and enable us to perform a manageable algorithm to determine the dimension of the isometry group. A flow diagram of this algorithm is also presented. 

In order to apply the method presented in the previous sections, we need to know if the spacetime admits a $R$-frame, and if so, determine it. At this point we can find different situations. For example, in Petrov-Bel types I, II and III, a frame can be algebraically obtained from the Weyl tensor. Although a method to determine this Weyl principal frame is known \cite{FMS-Weyl}, we will show in section \ref{sec-computing-H} that we can obtain the connection tensor without the explicit determination of the frame. Some cases where the $R$-frame depends on both the Weyl and Ricci tensors are also analyzed. 

Our above outlined approach implies that we get an IDEAL (Intrinsic, Deductive, Explicit and ALgorithmic) characterization of the geometries with a $R$-frame and admitting a G$_r$ group of isometries. Consequently, the algorithm can be easily implemented on the current tensor calculus packages. In section \ref{sec-xAct} we illustrate this fact by considering two examples, the Petrov homogeneous vacuum solution \cite{Petrov-sol} and the Wils conformally flat pure radiation solution \cite{Wils}, which we implement on the {\it xAct Mathematica} suite of packages. 

Finally, in section \ref{sec-discussion} we comment about our results, and we summarize some related topics and applications which are the subject of further work: the obtention of the causal character of the group orbits, the determination of the specific algebra type when a G$_2$, G$_3$ or G$_4$ exist, and the IDEAL characterization of the spatially homogeneous cosmologies.


\section{Isometries in Riemannian spaces with a Riemann-frame} 
\label{sec-Riemann-frame}

Most of the results that we obtain in this section can be inferred from the previous literature on spacetime invariants and their usefulness in the equivalence problems (see \cite{kramer, MacCallum-2015, tomoda} and the other references quoted in the introduction). Nevertheless, we prefer to present a quite self-contained reasoning that might be easier for the reader. 

Let us consider an $n$-dimensional Riemannian space where a $R$-frame $\{e_a\}_{a=1}^n$ exists. In what follows, we will use Latin indexes to count the vectors of the frame, and to indicate the components of a tensor in this non-holonomic frame, while we will use Greek indexes to indicate components in a coordinate frame. Every Killing vector $\xi$ leaves the $R$-frame invariant, ${\cal L}_\xi e_a =0$, and consequently,
\be \label{L-xi}
i (e_a) \nabla \xi = i(\xi) \nabla e_{a} \, ,
\ee
where $i(\xi)t$ represents the interior product of a vector field $\xi$ with a $p$-tensor $t$.

Thus, at every point, the components of the Killing two-form $\nabla \xi$ are determined by the components of the Killing vector $\xi$. Consequently, we have a known result: a Riemannian space that admits a $R$-frame has a trivial isotropy group and then, the dimension $r$ of the isometry group coincides with the dimension $s$ of its orbits. 

Note that tensors of the form $E_{\bar{q}} = e_{a_1}\! \otimes\!...\!\otimes e_{a_q}$, $\{e_{a}\}$ being a $R$-frame, which define a basis of the tensorial algebra, are also invariant for any Killing vector $\xi$, ${\cal L}_\xi E_{\bar{q}}=0$. Hereon, $\bar{q}$ denotes a multi-index, $\bar{q} = a_1 ... a_q$. With this notation, the $p$-covariant derivative of the Riemann tensor can be written as
\be \label{nablaR}
\nabla^p R= \kappa^{\bar{q}} E_{\bar{q}} \, , \qquad q=p+4 \, , \quad p \geq 0 \, .
\ee
Therefore, the integrability conditions of the Killing equation take the expression:
\be \label{kic-a}
\hspace{-20mm} 0= {\cal L}_\xi \nabla^p R= (\xi, \dif \kappa^{\bar{q}}) E_{\bar{q}} + \kappa^{\bar{q}}{\cal L}_\xi E_{\bar{q}} = (\xi, \dif \kappa^{\bar{q}}) E_{\bar{q}} +  \kappa^{\bar{q}} \xi^{\lambda} \nabla_{\lambda} E_{\bar{q}} - \displaystyle \kappa^{\bar{q}} F_{\!\bar{q}\,\mu}^{ \lambda} \nabla_{\lambda} \xi^\mu  \, ,
\ee
where the matrix of the coefficients of $\nabla \xi$, $F_{\!\bar{q}\,\mu}^{ \lambda}$, is an algebraic concomitant of the frame $\{e_{a}\}$. Since the isotropy group is trivial, and as a consequence of a result by Defrise \cite{Defrise}, this matrix has rank $n (n\!-\!1)/2$. Then, from (\ref{kic-a}), we can obtain $\nabla \xi$ in terms of $\xi$, an expression that must be compatible with (\ref{L-xi}) and that makes ${\cal L}_\xi E_{\bar{q}}$ vanish in (\ref{kic-a}). Consequently, these integrability conditions of the Killing equation become 
%
\be \label{kic-b}
(\xi, \dif \kappa^{\bar{q}}) =0   \, .
\ee
Note that the reasoning above is dimension and signature independent. Then, the paradigmatic theorem by Eisenhart \cite{eisenhart-33} implies: if a Riemannian space admits a $R$-frame, the dimension of the isometry group is $r = n-d$, where $d$ is the number of functionally independent Riemann invariant scalars. 

After this result, if we want to know the dimension of the isometry group of a Riemannian space with a $R$-frame, we must attend the following steps: (i) check that the spacetime indeed admits a $R$-frame and determine it from the Riemann tensor, (ii) make use of this $R$-frame to determine all the scalar invariants that can be obtained from the Riemann tensor and its covariant derivatives, (iii) study how many scalar invariants are independent.

The task of acquiring these steps is hard for several reasons. For instance, because we need to find a finite set of invariant scalars generating the whole set of scalars that can be obtained from the Riemann tensor and its derivatives. In the next subsection we show that working with the connection coefficients of the $R$-frame and their directional derivatives enables us to obtain such a set.


\subsection{A finite generating set of Riemann scalar invariants}
\label{subsec-dim-n}

Let $\{ e_{a} \}$ be a $R$-frame and $\{ \theta^{a} \}$ its algebraic dual basis. The connection coefficients $\gamma^{c}_{ab}$ are defined, as usual, by the condition:
\begin{equation} \label{con-sym}
\nabla e_a = \gamma_{ab}^c \, \theta^b \otimes e_c \, .
\end{equation}
The components of the Riemann tensor in the frame $\{ e_{a} \}$, $R = {R^n}_{cab} \, e_n\!\otimes\! \theta^c\! \otimes\! \theta^a \! \otimes \! \theta^b$, depend on $\{ \gamma^{a}_{bc} \}$ and their first directional derivatives, $e_f (\gamma^a_{bc})$, as:
\be
{R^n}_{cab} = e_a(\gamma^n_{cb}) - e_b(\gamma^n_{ca}) +
\gamma^m_{cb} \gamma^n_{ma} - \gamma^m_{ca} \gamma^n_{mb} +
(\gamma^m_{ab}- \gamma^m_{ba}) \gamma^n_{cm} \, .
\ee
Consequently, all the scalar invariants that can be built with the Riemann tensor and its successive covariant derivatives depend on the connection coefficients and their successive directional derivatives along the frame. We denote $\Gamma_{\!(q)}$ the set of directional derivatives of order $q$, namely,
\be
\Gamma_{\!(q)} = \{ e_{a_1}   \cdots e_{a_q} (\gamma^a_{bc} ) \} \, ,  \qquad q \geq 0 \, .
\ee
So, $\Gamma_{\!(0)} = \{\gamma^a_{bc} \}$, and $\Gamma_{\!(1)} = \{ e_{d}  (\gamma^a_{bc} ) \}$. With this notation, the set of the Riemann scalar invariants is defined by:
\be
\displaystyle \Gamma = \bigcup_{q=0}^{\infty} \Gamma_{\!(q)} = \{\gamma_{\hat{0}}, \gamma_{\hat{1}}, ... , \gamma_{\hat{q}},...\} \, ,  \qquad \gamma_{\hat{q}} = e_{a_1}   \cdots e_{a_q} (\gamma^a_{bc} ) \, .
\ee
Note that $\hat{q}$ is a $(q\!+\!3)$-multi-index, and $\gamma_{\hat{q}}$ denotes each of the $n^{q+2}(n-1)/2$ elements of the set $\Gamma_{\!(q)}$. 

We know that the number $d$ of independent functions that can be generated determines the dimension $r$ of the isometry group, $r=n-d$. We discuss below that it is enough to reach a finite derivation order to determine $d$.

Let us suppose that all the scalars of $\Gamma_{\!(q)}$ depend on the scalars of the previous sets $\Gamma_{\!(0)}$,
$\cdots$ $\Gamma_{\!(q-1)}$. Thus, we have that $ \gamma_{\hat{q}} =  \gamma_{\hat{q}}(\gamma_{\hat{0}}, \cdots , \gamma_{q\!\widehat{\,-}\!1})$. Then, taking a new directional derivative we obtain:
\be \hspace{-10mm}
\gamma_{q\!\widehat{\,+}\!1} = e_a(\gamma_{\hat{q}}) = \sum_{k=0}^{q-1} \frac{ \partial \gamma_{\hat{q}}}{\partial \gamma_{\hat{k}}}\,  e_a(\gamma_{\hat{k}}) = \sum_{k=1}^{q} 
\frac{ \partial \gamma_{\hat{q}}}{\partial \gamma_{k\!\widehat{\,-}\!1}}
\, \gamma_{\hat{k}} = \gamma_{q\!\widehat{\,+}\!1}(\gamma_{\hat{0}}, \cdots , \gamma_{q\!\widehat{\,-}\!1}) \, .
\ee
Therefore, we have shown the following. 
\begin{proposition}  \label{prop-generating}
Let $\gamma_{ab}^c$ be the connection coefficients associated with a $R$-frame $\{ e_{a} \}$.  If all the scalars of $\Gamma_{\!(q)}$ depend on the scalars of the previous sets $\Gamma_{\!(0)}$,
$\cdots$ $\Gamma_{\!(q-1)}$, then the scalars of the successive sets $\Gamma_{\!(q+1)}$, $\Gamma_{\!(q+2)}$ $\cdots$ also depend on them.
\end{proposition}
Thus, we have that if the directional derivatives of order $q$ of the connection coefficients depend on those of lower order, then those of order greater than $q$ also depend on those of lower order. In other words, the orders at which invariants appear cannot skip any value of $q$. Consequently, since $n$ is the maximum number of independent scalars, 
\be
\displaystyle \tilde{\Gamma} = \bigcup_{q=0}^{n-1} \Gamma_{\!(q)} = \{\gamma_{\hat{0}}, \gamma_{\hat{1}}, ... , \gamma_{n\!\widehat{\,-}\!1}\} \, ,  \qquad \gamma_{\hat{q}} = e_{a_1}   \cdots e_{a_q} (\gamma^a_{bc} ) \, ,
\ee
is a finite set generating all the Riemann invariant scalars. This fact enables us to perform a finite method to determine the dimension of the admitted group of isometries. Below, we specify it for the four-dimensional case. 


\subsection{The four-dimensional case}
\label{subsec-dim-4}

The results in this subsection and in the following section can be generalized for a generic $n$-dimensional Riemannian space (see subsection \ref{subsec-generalitzacio}). However, we focus on a Lorentzian four-dimensional case for its interest in General Relativity and because the results in subsequent sections are based on the properties of the Ricci and Weyl tensors of a four-dimensional space-time. 

In this four-dimensional case, we have $\tilde{\Gamma} = \{\gamma_{\hat{0}}, \gamma_{\hat{1}}, \gamma_{\hat{2}}, \gamma_{\hat{3}}\}$, and we analyze the functional constrains on the scalar invariants $\gamma_{\hat{q}}$ when an $r$-dimensional group of isometries G$_r$ exists.
%
\subsubsection*{Group {\rm G}$_4$} 
\label{subsubsec-G4}

This homogeneous case is characterized by the fact that all the invariants are constant. But, as a consequence of the above proposition \ref{prop-generating}, we only need to impose that the elements $\gamma_{\hat{0}} \in \Gamma_{\!(0)}$ are constant scalars:
\be \label{G4-gamma}
\hspace{-21.0mm} {\rm G}_4: \qquad \qquad  \qquad \dif \gamma_{\hat{0}} = 0 \, .
\ee
\subsubsection*{Group {\rm G}$_3$} 
\label{subsubsec-G3}

Now, the scalar invariants of $\Gamma_{\!(0)}$ must depend on a sole function, say $x$, $\gamma_{\hat{0}} = \gamma_{\hat{0}}(x)$, and the other elements of $\tilde{\Gamma}$ must also depend on $x$, $\gamma_{\hat{q}} = \gamma_{\hat{q}}(x)$. But,  as a consequence of proposition \ref{prop-generating}, we only need to impose this condition on the elements of $\Gamma_{\!(1)}$. Thus, the existence of a G$_3$ is characterized by conditions 
\be \label{G3-gamma}
\hspace{-20.0mm} {\rm G}_3: \qquad \qquad \{\dif \gamma_{\hat{0}}\} \not= \{0\} \, , \qquad \dif \gamma_{\hat{0}} \wedge \dif \gamma_{\hat{0}'} = 0 \,  , \qquad \dif \gamma_{\hat{0}} \wedge \dif \gamma_{\hat{1}} = 0 \, .
\ee

\subsubsection*{Group {\rm G}$_2$} 
\label{subsubsec-G2}

Now, the scalar invariants of  $\tilde{\Gamma}$ must generate two independent functions, say $x$ and $y$. In this case we have two possible situations. If $\gamma_{\hat{0}} = \gamma_{\hat{0}}(x, y)$ (case G$_{2a}$), the elements of $\Gamma_{\!(1)}$ must fulfill $\gamma_{\hat{1}} = \gamma_{\hat{1}}(x, y)$, and then this property holds for the other elements of $\tilde{\Gamma}$ as a consequence of proposition \ref{prop-generating}. If $\gamma_{\hat{0}} = \gamma_{\hat{0}}(x)$ and $\gamma_{\hat{1}} = \gamma_{\hat{1}}(x, y)$ (case G$_{2b}$), the elements of $\Gamma_{\!(2)}$ must fulfill $\gamma_{\hat{2}} = \gamma_{\hat{2}}(x, y)$, and then this property holds for the other elements of $\tilde{\Gamma}$ as a consequence of proposition \ref{prop-generating}. Thus, the existence of a G$_2$ is characterized by one of the following two sets of conditions: 
\begin{equation} \label{G2-gamma-a}
\hspace{-23mm} {\rm G}_{2a}: \qquad \quad \{\dif \gamma_{\hat{0}}\! \wedge\! \dif \gamma_{\hat{0}'} \} \not= \{0\}  , \quad \dif \gamma_{\hat{0}}\! \wedge\!  \dif \gamma_{\hat{0}'} \! \wedge\!  \dif \gamma_{\hat{0}''}  = 0   , \quad \dif \gamma_{\hat{0}} \! \wedge\!  \dif \gamma_{\hat{0}'} \! \wedge\!  \dif \gamma_{\hat{1}}  = 0  , 
\ee
\be \label{G2-gamma-b}
\hspace{-23mm} {\rm G}_{2b}: \ \ \dif \gamma_{\hat{0}}\! \wedge\! \dif \gamma_{\hat{0}'} = 0  , \ \ \{\dif \gamma_{\hat{0}}\! \wedge\! \dif \gamma_{\hat{1}} \} \not= \{0\} , \ \ \dif \gamma_{\hat{0}}\! \wedge\!  \dif \gamma_{\hat{1}} \! \wedge\!  \dif \gamma_{\hat{1}'}  =  \dif \gamma_{\hat{0}} \! \wedge\!  \dif \gamma_{\hat{1}} \! \wedge\!  \dif \gamma_{\hat{2}}  = 0 . 
\ee
%

\subsubsection*{Group {\rm G}$_1$} 
\label{subsubsec-G1}

Now, the scalar invariants of  $\tilde{\Gamma}$ must generate three independent functions, say $x$, $y$ and $z$. In this case we have four possible situations. If $\gamma_{\hat{0}} = \gamma_{\hat{0}}(x, y, z)$ (case G$_{1a}$), the elements of $\Gamma_{\!(1)}$ must fulfill $\gamma_{\hat{1}} = \gamma_{\hat{1}}(x, y, z)$, and then this property holds for the other elements of $\tilde{\Gamma}$ as a consequence of proposition \ref{prop-generating}. If $\gamma_{\hat{0}} = \gamma_{\hat{0}}(x, y)$ and $\gamma_{\hat{1}} = \gamma_{\hat{1}}(x, y, z)$ (case G$_{1b}$), the elements of $\Gamma_{\!(2)}$ must fulfill $\gamma_{\hat{2}} = \gamma_{\hat{2}}(x, y, z)$, and then this property holds for the other elements of $\tilde{\Gamma}$ as a consequence of proposition \ref{prop-generating}. If $\gamma_{\hat{0}} = \gamma_{\hat{0}}(x)$ and $\gamma_{\hat{1}} = \gamma_{\hat{1}}(x, y, z)$ (case G$_{1c}$), the elements of $\Gamma_{\!(2)}$ must fulfill $\gamma_{\hat{2}} = \gamma_{\hat{2}}(x, y, z)$, and then this property holds for the other elements of $\tilde{\Gamma}$ as a consequence of proposition \ref{prop-generating}. Finally, if $\gamma_{\hat{0}} = \gamma_{\hat{0}}(x)$, $\gamma_{\hat{1}} = \gamma_{\hat{1}}(x, y)$ and $\gamma_{\hat{2}} = \gamma_{\hat{2}}(x, y, z)$ (case G$_{1d}$), the elements of $\Gamma_{\!(3)}$ must fulfill $\gamma_{\hat{3}} = \gamma_{\hat{3}}(x, y, z)$. Thus, the existence of a G$_1$ is characterized by one of the following four sets of conditions: 
\begin{equation} \label{G1-gamma-a}
\hspace{-25.0mm} {\rm G}_{1a}: \ \  \{ \dif \gamma_{\hat{0}}\! \wedge\!  \dif \gamma_{\hat{0}'} \! \wedge\!  \dif \gamma_{\hat{0}''}\! \}\! \not=\! \{0\}  , \ \, \dif \gamma_{\hat{0}}\! \wedge\!  \dif \gamma_{\hat{0}'} \! \wedge\!  \dif \gamma_{\hat{0}''}  \! \wedge\!  \dif \gamma_{\hat{0}'''}\! =\! \dif \gamma_{\hat{0}}\! \wedge\!  \dif \gamma_{\hat{0}'} \! \wedge\!  \dif \gamma_{\hat{0}''}  \! \wedge\!  \dif \gamma_{\hat{1}} \!=\! 0  , 
\end{equation}
\begin{equation}  \label{G1-gamma-b}
\hspace{-25mm} {\rm G}_{1b}:  \ \ \left\lbrace
\begin{array}{c}
\! \! \{\dif \gamma_{\hat{0}}\! \wedge\! \dif \gamma_{\hat{0}'} \} \not= \{0\}  , \quad \dif \gamma_{\hat{0}}\! \wedge\!  \dif \gamma_{\hat{0}'} \! \wedge\!  \dif \gamma_{\hat{0}"}  = 0   , \quad \{\dif \gamma_{\hat{0}} \! \wedge\!  \dif \gamma_{\hat{0}'} \! \wedge\!  \dif \gamma_{\hat{1}}\}  \not= 0,   \\[2mm]
\! \! \dif \gamma_{\hat{0}}\! \wedge\!  \dif \gamma_{\hat{0}'} \! \wedge\!  \dif \gamma_{\hat{1}}  \! \wedge\!  \dif \gamma_{\hat{1}'}\! =\! 0 , \qquad \dif \gamma_{\hat{0}}\! \wedge\!  \dif \gamma_{\hat{0}'} \! \wedge\!  \dif \gamma_{\hat{1}}  \! \wedge\!  \dif \gamma_{\hat{2}} \!=\! 0  ,
\end{array}
\right.
\end{equation}

\begin{equation} \label{G1-gamma-c}
\hspace{-25mm} {\rm G}_{1c}:  
\ \ \left\{
\begin{array}{c}
\!\! \dif \gamma_{\hat{0}}\! \wedge\! \dif \gamma_{\hat{0}'}  = 0  ,\ \ \quad \{\dif \gamma_{\hat{0}} \! \wedge\!  \dif \gamma_{\hat{1}} \! \wedge\!  \dif \gamma_{\hat{1}'}\} \not= 0,   \\[2mm]
\! \dif \gamma_{\hat{0}}\! \wedge\!  \dif \gamma_{\hat{1}} \! \wedge\!  \dif \gamma_{\hat{1}'}  \! \wedge\!  \dif \gamma_{\hat{1}''}\! =\! 0 , \qquad \dif \gamma_{\hat{0}}\! \wedge\!  \dif \gamma_{\hat{1}} \! \wedge\!  \dif \gamma_{\hat{1}'}  \! \wedge\!  \dif \gamma_{\hat{2}} \!=\! 0  ,
\end{array}
\right.
\end{equation}
\begin{equation}  \label{G1-gamma-d}
\hspace{-25mm} {\rm G}_{1d}:  \ \ \left\{
\begin{array}{c}
\!\! \dif \gamma_{\hat{0}}\! \wedge\! \dif \gamma_{\hat{0}'}  = 0  ,\ \ \quad \{\dif \gamma_{\hat{0}}\! \wedge\!  \dif \gamma_{\hat{1}}\}  \not= 0   ,\ \ \quad \dif \gamma_{\hat{0}} \! \wedge\!  \dif \gamma_{\hat{1}} \! \wedge\!  \dif \gamma_{\hat{1}'} = 0,   \\[2mm]
\!\!\{\dif \gamma_{\hat{0}}\! \wedge\!  \dif \gamma_{\hat{1}} \! \wedge\!  \dif \gamma_{\hat{2}} \}\! \not=\! \{0\}  , \quad \dif \gamma_{\hat{0}}\! \wedge\!  \dif \gamma_{\hat{1}} \! \wedge\!  \dif \gamma_{\hat{2}}  \! \wedge\!  \dif \gamma_{\hat{2}'}\! =\! \dif \gamma_{\hat{0}}\! \wedge\!  \dif \gamma_{\hat{1}} \! \wedge\!  \dif \gamma_{\hat{2}}  \! \wedge\!  \dif \gamma_{\hat{3}} \!=\! 0  .
\end{array}
\right.
\end{equation}

\subsubsection*{No isometries} 
\label{subsubsec-G0}

When none of the conditions G$_1$, G$_2$, G$_3$ or G$_4$ hold, it follows that the spacetime admits no isometries.


\section{Dimension of the isometry  group in terms of the connection tensor} 
\label{sec-dimension}

In the four-dimensional case, we have 24 connection coefficients  $\gamma_{\hat{0}} \equiv \gamma_{ab}^c \in \Gamma_{(0)}$, and  $24 \cdot 4^q$ directional $q$-derivatives $\gamma_{\hat{q}} \in \Gamma_{(q)}$. Note that conditions (\ref{G4-gamma}) characterizing the existence of a G$_4$ involve 24 one-forms, and the characterization (\ref{G3-gamma}) of a G$_3$ also imposes the nullity of $2580$ two-forms. The number of conditions characterizing a G$_2$ or a G$_1$ is even bigger, and this fact makes implementing a manageable algorithm to determine the dimension of the isometry group hard. The connection tensor that we introduce below enables us to gather in a few differential concomitants all the differential scalars $\gamma_{\hat{q}}$ and thus, it facilitates the performance of this algorithm.

\subsection{The connection tensor and its differential concomitants}
\label{subsec-connection}

From now on, we consider a spacetime with metric $g$ of signature $\{-+++\}$ and metric volume element $\eta$. If a $R$-frame exists, we can always orthonormalize it and then work with such oriented orthonormal $R$-frame $\{ e_{a} \}$, $\eta(e_0, e_1, e_2, e_3) = 1$. Let $\{ \theta^{a} \}$ be its algebraic dual basis. We can collect the connection  coefficients in the connection tensor $H$ defined as
\begin{equation} \label{defH}
H= \gamma^{c}_{a b} \, \theta^{b} \otimes \theta^a \otimes e_{c} .
\end{equation}

We shall denote with the same symbol a tensor and the metric  equivalent
tensors that follow by raising and lowering indexes with $g$. In
that sense, the tensor that results from $H$ when raising the
second index can be obtained from the frame $\{ e_{a} \}$ as:
\begin{equation} \label{H-wedge}
H = - \frac12 \tilde{\eta}^{a b}\nabla e_{a} \bar{\wedge} e_b  \, , 
\end{equation}
where $\tilde{\eta}^{ab}$ is the signature symbol, $\tilde{\eta}^{ab}= diag(-1,1,1,1)$, and where, for a vector $w$ and a two-tensor $A$, $(A \bar{\wedge} w)_{\alpha \beta \gamma} = A_{\alpha \beta} w_\gamma \!- \! A_{\alpha \gamma} w_\beta$.

The tensor $H$ provides the covariant derivative of the vectors $e_{a}$  by contracting the second
index of $H$ with $e_{a}$, that is,
\begin{equation} \label{nabla-e}
\left( \nabla e_{a} \right)_{\lambda \mu} =H_{\lambda \nu \mu} (e_{a})^\nu .
\end{equation}
If $\xi$ is a Killing field that leaves the frame $\{ e_a \}$ invariant, we have that equation (\ref{L-xi}) holds for every $e_a$. Then, if we use (\ref{nabla-e}) to replace $\nabla e_a$, we obtain:
\begin{equation} \label{nabla-killing}
\nabla \xi = i(\xi) H \, . 
\end{equation}
Conversely, if $\xi$ is a field that satisfies equation above, then $\nabla \xi$ is antisymmetric and consequently it is a Killing field. Thus, we have:
\begin{proposition}
If $H$ is the   connection tensor associated with a $R$-frame $\{ e_{a} \}$, then $\xi$ is a Killing field if, and only if, {\em (\ref{nabla-killing})} holds.
\end{proposition}
The connection tensor contains the 24 connection coefficients $\gamma^c_{a b}$ associated with the frame $\{e_{a} \}$. When the frame is invariant, the tensor $H$ is also an invariant tensor. Then, if $\xi$ is a Killing field, we have that
the Lie derivative of $H$ vanishes, ${\cal L}_{\xi} H=0$. Computing this derivative and using (\ref{nabla-killing}) to replace the covariant derivative of $\xi$, we obtain that, for every Killing field $\xi$, $i(\xi)C^{[1]} =0$, where $C^{[1]} \equiv C^{[1]}(H)$ is the first-order differential concomitant of $H$ defined as:
\begin{equation} \label{c1}
 C^{[1]}_{{\alpha} \mu \nu \sigma} \equiv \nabla_{\alpha} H_{\mu\nu \sigma} + {H_{\alpha \mu}}^\rho H_{\rho \nu \sigma}  +
{H_{{\alpha} \nu}}^\rho  H_{\mu \rho \sigma} + {H_{{\alpha} \sigma}}^\rho  H_{\mu \nu \rho} \,  .
\end{equation}
If we take into account the definition (\ref{defH}) of $H$, we get that $C^{[1]}$ contains the first derivatives of the
connection coefficients:
\be \label{c1-b}
C^{[1]} = e_d (\gamma^a_{b c}) \ \theta^d \otimes \theta^c \otimes
\theta^b \otimes e_a =  \dif \gamma^a_{b c}  \otimes \theta^c \otimes
\theta^b \otimes e_a \, .
\ee
Similarly, the second derivatives $e_{a_2} e_{a_1} (\gamma^{d}_{b c})$ of the connection coefficients
can be collected in a tensor $C^{[2]} \equiv C^{[2]}(H)$ defined as:
\begin{equation} \label{c2}
\hspace*{-2.5cm} C^{[2]}_{\alpha_2 \alpha_1 \mu \nu \sigma} \!=\!
\nabla_{\alpha_2} C^{[1]}_{\alpha_1 \mu \nu \sigma} \!+\! {H_{\alpha_2
\alpha_1}}^{\rho} C^{[1]}_{\rho \mu \nu \sigma}\! +\! {H_{\alpha_2
\mu}}^{\rho} C^{[1]}_{\alpha_1 \rho  \nu \sigma}\! +\! {H_{\alpha_2
\nu}}^{\rho} C^{[1]}_{\alpha_1  \mu \rho  \sigma}\! +\! {H_{\alpha_2
\sigma}}^{\rho} C^{[1]}_{\alpha_1 \mu  \nu \rho } \, .
\end{equation}
And also, we can define $C^{[q]} \equiv C^{[q]}(H)$ as
\begin{equation} \label{cq}
\hspace*{-2.5cm} C^{[q]}_{\alpha_q \cdots \, \alpha_1 \mu \nu \sigma} =
 \nabla_{\alpha_q} C^{[q-\!1]}_{\alpha_{q-\!1} \cdots
\, \alpha_1 \mu \nu \sigma} + {H_{\alpha_{q}
\alpha_{q-\!1}}}^{\!\rho} C^{[q-\!1]}_{\rho \cdots \, \alpha_1 \mu \nu
\sigma} + \cdots + {H_{\alpha_q \sigma}}^{\!\rho}
C^{[q-\!1]}_{\alpha_{q-\!1} \cdots\, \alpha_1 \mu \nu \rho} \, .
\end{equation}
From definition (\ref{defH}) of $H$, we get that $C^{[q]}$ contains the $q$-derivatives of the
connection coefficients:
\be \label{cq-b}
\begin{array}{lll}
C^{[q]} &\! \!=\! \!& e_{a_q} \cdots e_{a_1} (\gamma^{c}_{a b}) \ \theta^{a_q} \otimes \cdots \otimes \theta^{a_1}
\otimes \theta^{b} \otimes \theta^{a} \otimes e_{c} = \\[2mm]
&\! \!=\!\! & \dif [e_{a_{q-\!1}} \cdots e_{a_1} (\gamma^{c}_{a
b})] \otimes \theta^{a_{q-\!1}} \otimes \cdots \otimes \theta^{a_1}
\otimes \theta^{b} \otimes \theta^{a} \otimes e_{c} \, .
\end{array}
\ee
%


\subsection{Dimension of the isometry  group} 
\label{subsec-dimensionTheorems}

In the characterization theorem that we present in this section, we use the differential concomitants $C^{[q]} \equiv C^{[q]}(H)$ as defined in (\ref{cq}). On the other hand, the second expression of $C^{[q]}$ given in equation (\ref{cq-b}) shows that we can use these concomitants of $H$ to express the invariant conditions G$_r$ given in subsection \ref{subsec-dim-4}. Thus, we can characterize the dimension of the isometry group in terms of the connection tensor.

Hereafter, we shall use $\eta(C^{[q]})$, $\eta(C^{[p]}\!, C^{[q]})$, $\eta(C^{[p]}\!,C^{[q]}\!,  C^{[r]})$, or $\eta(C^{[p]}\!,C^{[q]}\!,  C^{[r]}\!,  C^{[s]})$, to indicate the contraction of one, two, three or four  indexes of the volume element $\eta$ with the first index of the tensors $C^{[q]}$.

We can see that condition (\ref{G4-gamma}) characterizing a G$_4$ is equivalent to the nullity of $C^{[1]}$ (see expression (\ref{c1-b})). Consequently, we obtain:
\begin{theorem} \label{theo-G4}
Let us consider a spacetime metric admitting a $R$-frame with associated connection tensor $H$. Then, it
admits a {\em G}$_4$ group of isometries if, and only if,
\be 
\hspace{-21.0mm} {\rm G}_4: \qquad \qquad  \qquad C^{[1]} \equiv C^{[1]}(H) =0.
\ee
\end{theorem}


A non homogeneous spacetime admitting a $R$-frame is then identified by $C^{[1]}\neq 0$. Condition (\ref{G3-gamma}) characterizing a G$_3$ also imposes that all the one-forms $\{\gamma_{\hat{0}}, \gamma_{\hat{1}}\}$ are collinear. Then, taking into account the expression (\ref{cq-b}) of $C^{[q]}$ for $q=1,2$, these conditions can be stated as $\eta^{\alpha \beta \gamma \delta} C^{[1]}_{\alpha \lambda \mu \nu} C^{[1]}_{\beta \rho \sigma \tau} =0$, $\eta^{\alpha \beta \gamma \delta} C^{[1]}_{\alpha \lambda \mu \nu} C^{[2]}_{\beta \rho \sigma \tau \omega} =0$. 
Thus, with the convention previously prescribed, we obtain the following statement:
\begin{theorem} \label{theo-G3}
Let us consider a spacetime metric admitting a $R$-frame with associated connection tensor $H$. Then, it
admits a {\em G}$_3$ group of isometries if, and only if,
\be 
\hspace{-21.0mm} {\rm G}_3: \qquad \qquad  C^{[1]}\neq 0, \quad \eta\,(C^{[1]}\!,C^{[1]}) =0, \quad \eta\,(C^{[1]}\!,C^{[2]} )=0 \, .
\ee
\end{theorem}


Similarly, taking into account the expressions (\ref{cq-b}) of $C^{[q]}$ for $q=1,2,3$, we can impose the two conditions (\ref{G2-gamma-a}) and (\ref{G2-gamma-b}) characterizing, respectively, a G$_{2a}$ and a G$_{2b}$, in terms of tensors of the form $\eta(C^{[q]}\!, C^{[r]})$, $\eta(C^{[p]}\!,C^{[q]}\!,  C^{[r]})$. Then, we arrive to the following result:
\begin{theorem} \label{theo-G2}
Let us consider a spacetime metric admitting a $R$-frame with associated connection tensor $H$. Then, it
admits a {\em G}$_2$ group of isometries if, and only if, one of the following conditions holds:
\be
\hspace{-25mm} {\rm G}_{2a}: \ \eta  ( C^{[1]}\!,C^{[1]} ) \neq 0, \quad \eta( C^{[1]}\!,C^{[1]}\!,C^{[1]})
=0, \quad \eta ( C^{[1]}\!,C^{[1]}\!,C^{[2]}) =0,
\ee
\be
\hspace{-25mm} {\rm G}_{2b}: \ \eta  ( C^{[1]}\!,C^{[1]} )\! =\! 0, \ \ \eta  ( C^{[1]}\!,C^{[2]} )\! \neq \!
0, \ \ \eta( C^{[1]}\!,C^{[2]}\!,C^{[2]}) \!=\!0, \ \ \eta ( C^{[1]}\!,C^{[2]}\!,C^{[3]})\! =\!0.
\ee
\end{theorem}


Again, taking into account the expressions (\ref{cq-b}) of $C^{[q]}$ for $q=1,2,3,4$, we can impose the four conditions (\ref{G1-gamma-a}), (\ref{G1-gamma-b}), (\ref{G1-gamma-c}) and (\ref{G1-gamma-d}) characterizing, respectively, a G$_{1a}$, a G$_{1b}$, a G$_{1c}$ and a G$_{1d}$, in terms of tensors of the form $\eta(C^{[q]}\!, C^{[r]})$, $\eta(C^{[p]}\!,C^{[q]}\!,  C^{[r]})$, $\eta(C^{[p]}\!, C^{[q]}\!,  C^{[r]}\!,  C^{[s]})$. Then, we obtain:
\begin{theorem} \label{theo-G1}
Let us consider a spacetime metric admitting a $R$-frame with associated connection tensor $H$. Then, it
admits a {\em G}$_1$ group of isometries if, and only if, one of the following conditions holds:
\be
\hspace{-25mm} {\rm G}_{1a}: \   \eta(C^{[1]}\!,C^{[1]}\!,C^{[1]}) \neq 0, \quad \eta(C^{[1]}\!,C^{[1]}\!,C^{[1]}\!,C^{[1]} ) = 0 , \quad \eta(C^{[1]}\!,C^{[1]}\!,C^{[1]}\!,C^{[2]}) = 0,
\ee
\be
\hspace{-25mm} {\rm G}_{1b}: \ \left\{ \begin{array}{l} \eta(C^{[1]}\!,C^{[1]} ) \neq 0, \quad \eta(C^{[1]}\!,C^{[1]}\!,C^{[1]}) = 0, \quad \eta(C^{[1]}\!,C^{[1]}\!,C^{[2]}) \neq 0 ,\\[2mm]
 \eta(C^{[1]}\!,C^{[1]}\!,C^{[2]}\!,C^{[2]}) = 0, \quad  \eta(C^{[1]}\!,C^{[1]}\!,C^{[2]}\!,C^{[3]}) = 0 ,
\end{array} \right.
\ee
\be
\hspace{-25mm} {\rm G}_{1c}: \ \left\{ \begin{array}{l}  \eta(C^{[1]}\!,C^{[1]} ) = 0, \quad \eta(C^{[1]}\!,C^{[2]}\!,C^{[2]}) \neq  0 ,
\\[2mm]
 \eta(C^{[1]}\!,C^{[2]}\!,C^{[2]}\!,C^{[2]}) = 0, \quad  \eta(C^{[1]}\!,C^{[2]}\!,C^{[2]}\!,C^{[3]}) = 0 ,
\end{array} \right.
\ee
\be
\hspace{-25mm} {\rm G}_{1d}: \ \left\{ \begin{array}{l}
  \eta\,(C^{[1]}\!,C^{[1]}  ) = 0, \quad  \eta(C^{[1]}\!,C^{[2]} ) \neq  0, \quad  \eta(C^{[1]}\!,C^{[2]}\!,C^{[2]} ) =
  0, \\[2mm]  \eta(C^{[1]}\!,C^{[2]}\!,C^{[3]}) \neq  0, \ \
   \eta(C^{[1]}\!,C^{[2]}\!,C^{[3]}\!,C^{[3]}) = 0, \ \  \eta(C^{[1]}\!,C^{[2]}\!,C^{[3]}\!,C^{[4]}) = 0 .
\end{array} \right.
\ee
\end{theorem}


Finally, when none of the conditions G$_r$ of the above theorems holds, the spacetime metric has no symmetries:
\begin{theorem} \label{theo-G0}
Let us consider a spacetime metric admitting a $R$-frame with associated connection tensor $H$. Then, there exists no Killing vector if, and only if, $H$  does not satisfy the conditions of any of the theorems  {\em \ref{theo-G4}, \ref{theo-G3}, \ref{theo-G2} and \ref{theo-G1}}.
\end{theorem}



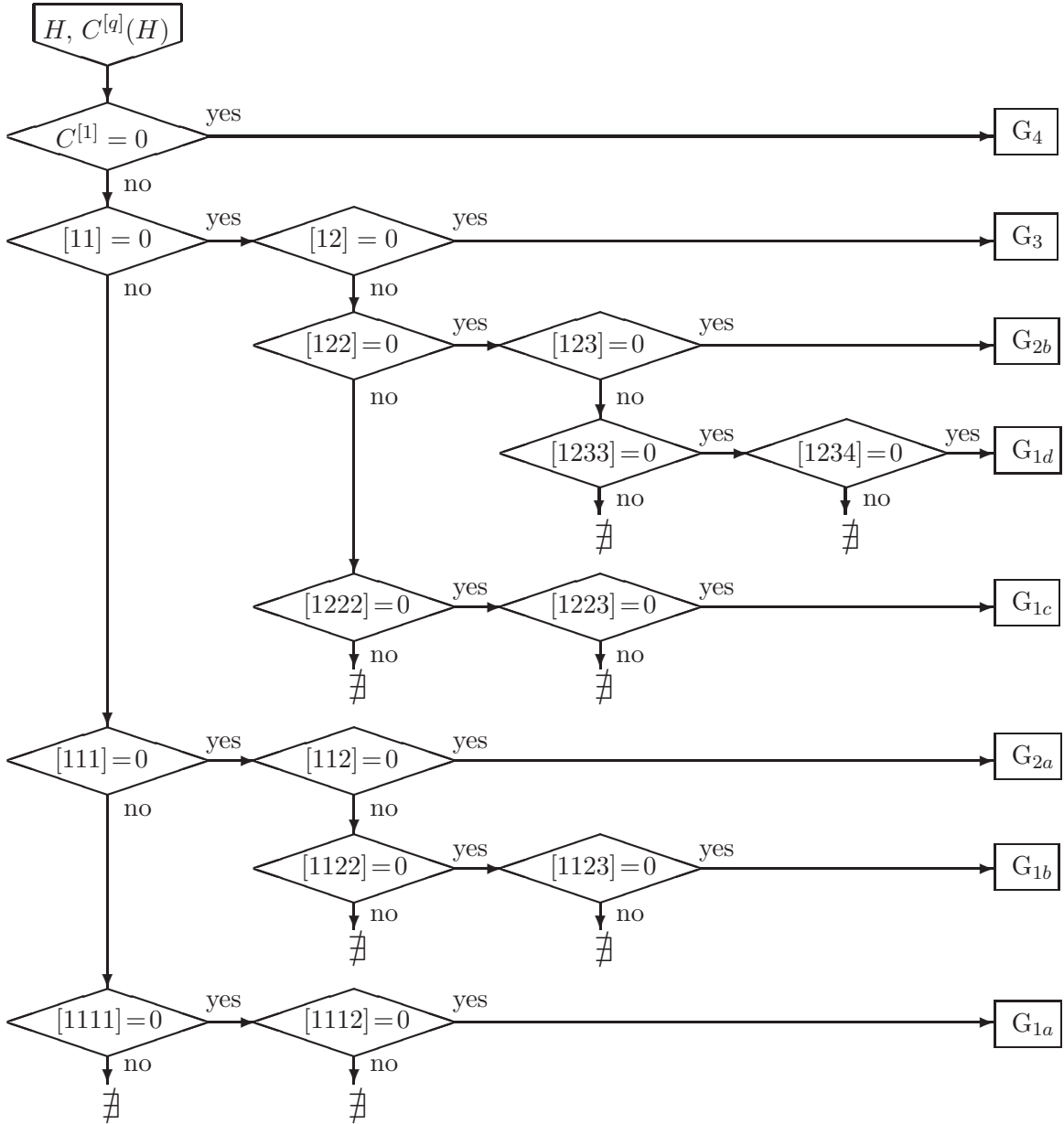
\begin{figure}[t]

\vspace*{7mm}
\hspace*{10mm} \setlength{\unitlength}{0.895cm} {\small \noindent
\begin{picture}(0,18)
\thicklines


\put(1.7,18){\line(-3,-1){1.2}}
 \put(-0.7,18){\line(3,-1){1.2}}
\put(1.7,18){\line(0,1){0.6}} \put(-0.7,18.6){\line(1,0){2.4}}
\put(-0.7,18.6){\line(0,-1){0.6}}

 \put(-0.55,18.05  ){$ H, \,  C^{[q]}(H)  $}
 

\put(2.1,16.7){yes}

 \put(2.1,15){yes}

 \put(6.1,15){yes}

 \put(6.1,13.3){yes}

 \put(10.1,13.3){yes}

 \put(10.1,11.5){yes}

\put(14.1,11.5){yes}

\put(2.1,6.5){yes}

\put(6.1,6.5){yes}

\put(6.1,4.75){yes}

\put(10.1,4.75){yes}

\put(6.1,2.3){yes}

\put(2.1,2.3){yes}

\put(6.1,9){yes}

\put(10.1,9){yes}


\put(0.76,15.56){no}

\put(0.76,13.86){no}

\put(4.76,13.86){no}

\put(4.76,12.1){no}

\put(8.76,12.1){no}

\put(8.76,10.45){no}

\put(12.76,10.45){no}

\put(8.76,7.9){no}

\put(4.76,7.9){no}

\put(4.76,5.4){no}

\put(0.76,5.4){no}

\put(4.76,3.7){no} \put(8.76,3.7){no}

\put(0.76,1.25){no}\put(4.76,1.25){no}


\put(0.5,17.60){\vector(0,-1){0.6 }}

\put(0.5,15.9){\vector(0,-1){0.6 }}

\put(0.5,14.2){\vector(0,-1){7.35 }}

\put(0.5,5.75){\vector(0,-1){3.15 }}

\put(4.5,14.2){\vector(0,-1){0.6 }}

\put(4.5,12.5){\vector(0,-1){3.15 }}

\put(4.5,5.75){\vector(0,-1){0.65 }}

\put(8.5,12.5){\vector(0,-1){0.65 }}


\put(2.1,16.45){\vector(1,0){12.8}}

\put(2.1,14.75){\vector(1,0){0.8}}

\put(6.1,14.75){\vector(1,0){8.8}}

\put(6.1,13.05){\vector(1,0){0.8}}

\put(10.1,13.05){\vector(1,0){4.8}}

\put(10.1,11.3){\vector(1,0){0.8}}

\put(14.1,11.3){\vector(1,0){0.8}}

\put(6.1,8.8){\vector(1,0){0.8}}

\put(10.1,8.8){\vector(1,0){4.8}}

\put(2.1,6.3){\vector(1,0){0.8}}

\put(6.1,6.3){\vector(1,0){8.8}}

\put(6.1,4.55){\vector(1,0){0.8}}

\put(10.1,4.55){\vector(1,0){4.8}}

\put(2.1,2.05){\vector(1,0){0.8}}

\put(6.1,2.05){\vector(1,0){8.8}}


\put(0.5,17){\line(-3,-1){1.62}}  \put(0.5,17){\line(3,-1){1.62}}
\put(0.5,15.9){\line(3,1){1.62}} \put(0.5,15.9){\line(-3,1){1.62}}
\put(-0.34,16.27){$C^{[1]} =0$}


\put(14.9,16.4){\framebox{\
G$_4$\,\hspace*{-0.4cm}$\phantom{{A^B_C}}\!\!$}}


\put(0.5,15.3){\line(-3,-1){1.62}} \put(0.5,15.3){\line(3,-1){1.62}}
\put(0.5,14.2){\line(3,1){1.62}} \put(0.5,14.2){\line(-3,1){1.62}}
\put(-0.24,14.65){$[11]=0$}

\put(4.5,15.3){\line(-3,-1){1.62}} \put(4.5,15.3){\line(3,-1){1.62}}
\put(4.5,14.2){\line(3,1){1.62}} \put(4.5,14.2){\line(-3,1){1.62}}
\put(3.76,14.65){$[12]=0$}

\put(14.9,14.7){\framebox{\
G$_3$\,\hspace*{-0.4cm}$\phantom{{A^B_C}}\!\!$}}


\put(4.5,13.6){\line(-3,-1){1.62}} \put(4.5,13.6){\line(3,-1){1.62}}
\put(4.5,12.5){\line(3,1){1.62}} \put(4.5,12.5){\line(-3,1){1.62}}
\put(3.7,12.95){$[122]\!=\!0$}

\put(8.5,13.6){\line(-3,-1){1.62}} \put(8.5,13.6){\line(3,-1){1.62}}
\put(8.5,12.5){\line(3,1){1.62}} \put(8.5,12.5){\line(-3,1){1.62}}
\put(7.7,12.95){$[123]\!=\!0$}

\put(14.9,13){\framebox{\
G$_{2b}$\,\hspace*{-0.5cm}$\phantom{{A^B_C}}\!\!$}}


\put(8.5,11.85){\line(-3,-1){1.62}}
\put(8.5,11.85){\line(3,-1){1.62}} \put(8.5,10.75){\line(3,1){1.62}}
\put(8.5,10.75){\line(-3,1){1.62}} \put(7.68,11.17){$[1233]\!=\!0$}

\put(8.4,9.75){{\large $\nexists$}}
\put(8.48,10.75){\vector(0,-1){0.45}}

\put(12.5,11.85){\line(-3,-1){1.62}}
\put(12.5,11.85){\line(3,-1){1.62}}
\put(12.5,10.75){\line(3,1){1.62}}
\put(12.5,10.75){\line(-3,1){1.62}}
\put(11.68,11.17){$[1234]\!=\!0$}

\put(12.4,9.75){{\large $\nexists$}}
\put(12.48,10.75){\vector(0,-1){0.45}}

\put(14.9,11.22){\framebox{\
G$_{1d}$\,\hspace*{-0.5cm}$\phantom{{A^B_C}}\!\!$}}


\put(4.5,9.35){\line(-3,-1){1.62}} \put(4.5,9.35){\line(3,-1){1.62}}
\put(4.5,8.25){\line(3,1){1.62}} \put(4.5,8.25){\line(-3,1){1.62}}
\put(3.68,8.7){$[1222]\!=\!0$}

\put(8.5,9.35){\line(-3,-1){1.62}} \put(8.5,9.35){\line(3,-1){1.62}}
\put(8.5,8.25){\line(3,1){1.62}} \put(8.5,8.25){\line(-3,1){1.62}}
\put(7.68,8.7){$[1223]\!=\!0$}

\put(14.9,8.75){\framebox{\
G$_{1c}$\,\hspace*{-0.5cm}$\phantom{{A^B_C}}\!\!$}}

\put(4.5,8.25){\vector(0,-1){0.45}} \put(4.4,7.3){{\large
$\nexists$}}

 \put(8.4,7.3){{\large $\nexists$}}

\put(8.5,8.25){\vector(0,-1){0.45}}


\put(0.5,6.85){\line(-3,-1){1.62}} \put(0.5,6.85){\line(3,-1){1.62}}
\put(0.5,5.75){\line(3,1){1.62}} \put(0.5,5.75){\line(-3,1){1.62}}
\put(-0.36,6.2){$[111]\!=\!0$}

\put(4.5,6.85){\line(-3,-1){1.62}} \put(4.5,6.85){\line(3,-1){1.62}}
\put(4.5,5.75){\line(3,1){1.62}} \put(4.5,5.75){\line(-3,1){1.62}}
\put(3.7,6.2){$[112]\!=\!0$}

\put(14.9,6.25){\framebox{\
G$_{2a}$\,\hspace*{-0.5cm}$\phantom{{A^B_C}}\!\!$}}


\put(4.5,5.1){\line(-3,-1){1.62}} \put(4.5,5.1){\line(3,-1){1.62}}
\put(4.5,4){\line(3,1){1.62}} \put(4.5,4){\line(-3,1){1.62}}
\put(3.64,4.45){$[1122]\!=\!0$}

\put(8.5,5.1){\line(-3,-1){1.62}} \put(8.5,5.1){\line(3,-1){1.62}}
\put(8.5,4){\line(3,1){1.62}} \put(8.5,4){\line(-3,1){1.62}}
\put(7.68,4.45){$[1123]\!=\!0$}

\put(14.9,4.45){\framebox{\
G$_{1b}$\,\hspace*{-0.5cm}$\phantom{{A^B_C}}\!\!$}}

\put(8.4,3.05){{\large $\nexists$}}

\put(8.5,4){\vector(0,-1){0.45}}

\put(4.4,3.05){{\large $\nexists$}}

\put(4.5,4){\vector(0,-1){0.45}}


\put(0.5,2.6){\line(-3,-1){1.62}}  \put(0.5,2.6){\line(3,-1){1.62}}
\put(0.5,1.5){\line(3,1){1.62}} \put(0.5,1.5){\line(-3,1){1.62}}
\put(-0.34,1.95){$[1111]\!=\!0$}

\put(4.5,2.6){\line(-3,-1){1.62}}  \put(4.5,2.6){\line(3,-1){1.62}}
\put(4.5,1.5){\line(3,1){1.62}} \put(4.5,1.5){\line(-3,1){1.62}}
\put(3.68,1.95){$[1112]\!=\!0$}

\put(4.4,0.5){{\large $\nexists$}}

\put(4.5,1.5){\vector(0,-1){0.45}}

\put(0.4,0.5){{\large $\nexists$}}

\put(0.5,1.5){\vector(0,-1){0.45}}

 \put(14.9,1.9){\framebox{\
G$_{1a}$\,\hspace*{-0.5cm}$\phantom{{A^B_C}}\!\!$}}

\end{picture} }

\vspace{-5mm}
\caption{This flow diagram distinguishes the dimension of the groups of isometries of a spacetime admitting a Riemann-frame. We use the following notation: $[qr] \equiv \eta(C^{[q]}\!, C^{[r]})$, $[p q r] \equiv \eta(C^{[p]}\!, C^{[q]}\!,  C^{[r]})$, $[p q r s] \equiv \eta(C^{[p]}\!, C^{[q]}\!,  C^{[r]}\!,  C^{[s]})$, where these concomitants of $\eta$ and $C^{[p]}$ are defined in subsection \ref{subsec-dimensionTheorems}. }
\label{figure-1}
\end{figure}


\subsection{Caracterization algorithm}
\label{algorithm}

The theorems stated above enable us to carry out a flowchart (figure \ref{figure-1}) that performs an algorithm providing
the dimension of the isometry group of a spacetime admitting a $R$-frame. This flow diagram uses as initial input data the connection tensor $H$ and its differential concomitants $C^{[q]}$, $q=1,2,3,4$, defined in subsection \ref{subsec-connection}. In order to improve the format of the flowchart we use the following notation: 
$[qr] \equiv \eta(C^{[q]}\!, C^{[r]})$, $[p q r] \equiv \eta(C^{[p]}\!, C^{[q]}\!,  C^{[r]})$, $[p q r s] \equiv \eta(C^{[p]}\!, C^{[q]}\!,  C^{[r]}\!,  C^{[s]})$. In the flowchart, the hori\-zon\-tal end arrows lead to the different G$_r$, and the vertical end arrows lead to no symmetries. Note that the four columns of diamond conditions correspond to the order of differentiation of these conditions.


\section{Computing the connection tensor. IDEAL characterization}
\label{sec-computing-H}

The previous section shows that we can determine the dimension of the isometry group of a spacetime with a Riemann-frame in terms of the connection tensor associated with it. If one knows explicitly the $R$-frame $\{ e_{a} \}$, expression (\ref{H-wedge}) can be used to compute $H$ and then the results of the previous section apply. 

But most of the time, the explicit determination of the frame is not necessary and $H$ can be obtained in terms of an
adequate invariant tensor related to the frame. This happens, for example, in dealing with the three-dimensional Riemannian and Lorentzian spaces, where the Ricci tensor or a differential concomitant of it can be used to obtain the connection tensor \cite{FS-K3,FS-G3,FS-L3}. 

In the four-dimensional Lorentzian case considered here, the existence of such a frame strongly relies on the geometric properties of the curvature tensor and, more specifically, on the geometric elements associated with the Weyl and the Ricci tensors. 

In spacetimes of Petrov-Bel types I, II or III, the Weyl tensor defines a principal frame. In \cite{FMS-Weyl} we have presented an algorithmic way to determine this Weyl frame. Nevertheless, we see in the following three subsections that the associated connection tensor can be determined without explicitly obtaining the frame.

In this section we use the notation adopted in \cite{FMS-Weyl} when getting certain concomitants of the Weyl tensor. The self-dual Weyl tensor is ${\cal W}= \frac{1}{2} (W - \ci *W)$, and ${\cal G} = \frac{1}{2} (G- \ci \eta)$ denotes the induced metric on the space of the self-dual bivectors, $G_{\alpha \beta \lambda \mu} = g_{\alpha \lambda} g_{\beta \mu} - g_{\alpha \mu} g_{\beta \lambda}$. We denote ${\cal W}^2$ the double bivector $({\cal W}^2)_{\alpha \beta \mu \nu} = \frac{1}{2} {\cal W}_{\alpha \beta \lambda \rho} {{\cal W}^{\lambda \rho}}_{\mu \nu}$, and $\Tr {\cal W} = \frac{1}{2}  {{\cal W}^{\mu \nu}}_{\mu \nu}$.


\subsection{Type I spacetimes}
\label{subsec-typeI}

In a type I spacetime, the self-dual Weyl tensor ${\cal W}$ is algebraically general and it diagonalizes in an orthonormal frame of bivectors $\{ {\cal U}_A \}$, $A=1,2,3$,  ${\cal W}= -  \sum \alpha_A \, {\cal U}_A \otimes {\cal U}_A$ \cite{FMS-Weyl}. This frame determines the Weyl principal frame of vectors $\{e_a\}$ (see \ref{sec-bivectors}).  Then, we can compute the self-dual connection tensor ${\cal H} = (H - \ci *H)/\sqrt{2}$ in a similar way to how we compute it in the three-dimensional Riemannian case \cite{FS-K3}. More specifically, we obtain:
\begin{proposition} 
\label{propo-typeI}
Let ${\cal W}$ be a type I Petrov-Bel Weyl tensor. The connection tensor $H$ associated with the Weyl principal frame can be obtained as
\be \label{H-Prop3}
H =\frac{1}{\sqrt{2}}({\cal H} + \bar{\cal H}) \, ,  \qquad {\cal H}_{\alpha   \mu \nu} \equiv \frac{1}{\sqrt{2}} {\cal X}_{\alpha \lambda \rho} {{\cal Y}^{\lambda \rho}}_{\mu \nu} \, ,
\ee
\be
{\cal X}_{\lambda \alpha \rho} \equiv \frac{1}{2} \nabla_{\lambda} {\cal W}_{\alpha \beta \mu \nu}  {{\cal W}^{\mu \nu \beta}}_{\rho}\, , \qquad {\cal Y} \equiv \frac{1}{\Delta} (3 a {\cal W}^2 + 6 b {\cal W} +
\frac{1}{2} a^2 {\cal G} ) \, ,
\ee
where $a\equiv \Tr {\cal W}^2$, $b\equiv \Tr {\cal W}^3$ and $\Delta \equiv a^3 - 6 b^2 \not= 0$.
\end{proposition}
The proof of the above proposition follows from the canonical expression of the self-dual Weyl tensor and expression (\ref{nablaU}). Indeed, a straightforward calculation leads to: 
\begin{eqnarray}
{\cal X}=  - \frac{1}{2} \sum_{A=1}^3 \alpha_{A} \dif  \alpha_A
\otimes g - \frac{\ci}{\sqrt{2}} \sum_{(ABC)} (\alpha_A - \alpha_B)^2
\Gamma_A^B \otimes {\cal U}_C \, , \\
 {\cal Y}= - \sum_{(ABC)} \frac{1}{(\alpha_A - \alpha_B)^2} {\cal
U}_C \otimes {\cal U}_C \, ,
\end{eqnarray}
where $(ABC)$ denotes cyclic permutation of $(1\,2\,3)$. Then, (\ref{H-Prop3}) follows taking into account (\ref{calH}).


\subsection{Type II spacetimes}
\label{subsec-typeII}

The self-dual Weyl tensor of a type II spacetime admits the canonical form \cite{FMS-Weyl} ${\cal W} = 3 \rho \, {\cal U} \otimes {\cal U} + \rho \, {\cal G} + {\cal L}_+ \otimes {\cal L}_+$. The eigen-bivectors ${\cal U}$ (unitary) and ${\cal L}_+$ (null) determine an oriented frame of bivectors $\{{\cal U}, {\cal L}_+, {\cal L}_-\}$, 
\be \label{L-}
\hspace{-15mm} {\cal L}_- = \frac{1}{2 ({\cal L}_+ , {\cal X})^2} \left[{\cal S}({\cal X}, {\cal X}) {\cal L}_+ - 2({\cal L}_+ , {\cal X}) {\cal S}({\cal X}) \right] ,  \qquad {\cal S} \equiv {\cal G} + {\cal U} \otimes {\cal U} \, ,
\ee
where ${\cal X}$ is a bivector such that $({\cal L}_+, {\cal X}) \neq 0$. This frame of bivectors determines a Weyl principal frame of vectors $\{e_a\}$. Moreover, some expressions in reference \cite{FMS-Weyl} provide $\cal U$ and $\cal L_+$ in terms of the Weyl tensor. Then, expression (\ref{SDnullct}) leads to: 
\begin{proposition} 
\label{propo-typeII}
Let ${\cal W}$ be a type II Petrov-Bel Weyl tensor. The connection tensor $H$ associated with the Weyl principal frame can be obtained as
\be \label{H-II-III}
\hspace{-15mm} H =\frac{1}{\sqrt{2}}({\cal H} + \bar{\cal H})  ,  \qquad {\cal H} \equiv - \frac{1}{\sqrt{2}} \left(  \nabla {\cal U} \cdot {\cal U} +  \nabla {\cal L}_+\! \cdot \! {\cal L}_- +  \nabla {\cal L}_- \! \cdot \! {\cal L}_+  \right)  ,
\ee
\be
\hspace{0mm}  {\cal U}= \frac{{\cal P}({\cal Y})}{\sqrt{-{\cal P}^2({\cal Y},{\cal Y})}},  \qquad  {\cal L}_+= \frac{{\cal Q}({\cal X})}{\sqrt{{\cal Q}({\cal X},{\cal X})}},  
\ee
with ${\cal L}_-$ given in {\em (\ref{L-})}, ${\cal P} \equiv ({\cal W}- \rho {\cal G})^2$ , ${\cal Q} \equiv \frac{1}{3 \rho} ({\cal W}- \rho {\cal G})({\cal W}+ 2 \rho {\cal G})$, $\rho \equiv -b/a$,  $a \equiv \Tr {\cal W}^2$ and $b \equiv \Tr {\cal W}^3$, and where ${\cal Y}$ and ${\cal X}$ are arbitrary bivectors such that  ${\cal P}({\cal Y})\not=0$ and ${\cal Q}({\cal X})\not=0$, respectively.
\end{proposition}
%


\subsection{Type III spacetimes}
\label{subsec-typeIII}

The self-dual Weyl tensor of a type III spacetime admits the canonical form \cite{FMS-Weyl} ${\cal W}= U \widetilde{\otimes} {\cal L}_+$. The null eigen-bivector ${\cal L}_+$ and the characteristic unitary bivector ${\cal U}$ determine an oriented frame of bivectors $\{{\cal U}, {\cal L}_+, {\cal L}_-\}$ as (\ref{L-}), which in turn determines a Weyl principal frame of vectors $\{e_a\}$. Moreover, some expressions in reference \cite{FMS-Weyl} provide $\cal U$ and $\cal L_+$ in terms of the Weyl tensor. Then, expression (\ref{SDnullct}) leads to: 
\begin{proposition} 
\label{propo-typeIII}
Let ${\cal W}$ be a type III Petrov-Bel Weyl tensor. The connection tensor $H$ associated with the Weyl principal frame can be obtained as {\em (\ref{H-II-III})}, where
\be
\hspace{-20mm}  {\cal L}_+ = - \frac{{\cal W}^2({\cal X})}{\sqrt{- {\cal W}^2({\cal X}, {\cal X})}}, \qquad  {\cal U}= \frac{1}{2 ({\cal L}_+ ,{\cal X})^2} \Big[ 2 ({\cal L}_+ , {\cal X}) {\cal W}({\cal X}) - {\cal W}({\cal
X}, {\cal X}) {\cal L}_+ \Big] ,
\ee
and ${\cal L}_-$ is given in {\em (\ref{L-})}, and where ${\cal X}$ is an arbitrary bivector such that  ${\cal W}^2({\cal X})\not=0$.
\end{proposition}
%


\subsection{When an invariant time-like vector $u$ and an orthogonal $2$-tensor $E$ are known}
\label{subsec-Eu}

There are quite a few situations where the orthonormal $R$-frame $\{u, e_A\}$, $A=1,2,3$, is defined by a unit time-like vector $u$ and the three eigenvectors of a symmetric 2-tensor $E$, which is orthogonal to $u$ and admits three different eigenvalues. 

For example, $u$ could be the unit velocity of a perfect fluid solution, or it could be defined by the gradient of a non-constant scalar invariant. On the other hand, $E$ could be the electric part of the Weyl tensor with respect to $u$, when the Weyl tensor is Petrov-Bel type N or type D (with $u$ outside of the principal plane); or also, $E$ could be the shear tensor of $u$. 


From the expression (\ref{H-wedge}) of the connection tensor in terms of the $R$-frame, a straightforward computation leads to the following result: 
\begin{proposition} \label{propo-uE}
Let $u$ be a unit time-like direction and $E$ a symmetric traceless {\em 2}-tensor with $\delta \equiv (\tr E^2)^3-6(\tr E^3)^2 \neq 0$ and $E(u)=0$. The connection tensor $H$ associated with the frame defined by $u$ and the eigenframe of $E$ can be obtained as:
\be
\hspace{-20mm} H= \nabla u \bar{\wedge} u - J  , \qquad J_{\alpha \gamma \lambda} \equiv \nabla_\alpha E^{\mu}_{\ \rho} \, E^{\rho \nu} \,  \epsilon_{\mu\nu \pi} K^{\pi \sigma} \epsilon_{\sigma \gamma \lambda} , \quad \epsilon_{\sigma \gamma \lambda} \equiv  \eta_{\sigma \gamma \lambda \delta} u^{\delta}  , 
\ee
\be
\hspace{-20mm}  K \equiv \frac{1}{\delta} [ 3 \alpha E^2 + 6 \beta E + \frac{1}{2} \alpha^2 (g\!+\! u \otimes u)] \, , \quad  \alpha \! \equiv \! \tr E^2, \ \ \beta\! \equiv \! \tr E^3 , \ \ \delta\! \equiv \! \alpha^3 - 6 \beta^2   .
\ee
\end{proposition}
%


\subsection{IDEAL characterization}
\label{subsec-IDEAL}

When the connection tensor $H$ defined by a $R$-frame is obtained as an explicit concomitant of the Riemann tensor, the theorems and the algorithm presented in section \ref{sec-dimension} become an IDEAL (Intrinsic, Deductive, Explicit and ALgorithmic) characterization of the geometries admitting a G$_r$ group of isometries. For example, for Petrov-Bel spacetimes I, II or III, we obtain:
\begin{corollary} \label{coro-IDEAL-Weyl}
A spacetime of Petrov-Bel type I (respectively, II or III) admits a group of isometries of dimension four, three, two, one, or no symmetries if, and only if, the connection tensor $H$ obtained in proposition {\em \ref{propo-typeI}} (respectively, {\em \ref{propo-typeII}} or {\em \ref{propo-typeIII}}) fulfills, respectively, the conditions of theorems {\em \ref{theo-G4}, \ref{theo-G3}, \ref{theo-G2}, \ref{theo-G1} or \ref{theo-G0}}.
\end{corollary}

In Petrov-Bel types O, N and D the algebraic structure of the Weyl tensor does not determine a Weyl-frame to compute $H$. Then, we need to obtain new invariant directions from the Ricci tensor or from the derivatives of the Weyl tensor. Elsewhere, we consider several of these circumstances in studying an invariant approach to the spatially-homogeneous cosmologies \cite{FS-SHC}. As an example, we present now the case of a type N perfect fluid solution. 

If $u$ is the unit velocity of a type N perfect fluid solution, the electric part $E$ of the Weyl tensor fulfills the conditions of proposition \ref{propo-uE}. Consequently, we can state:
\begin{corollary} \label{coro-IDEAL-N}
Let us consider a Petrov-Bel type N perfect fluid solution, $u$ its unit velocity, and $E$ the electric part of the Weyl tensor, $E_{\alpha \beta} = u^{\lambda} u^{\mu} W_{\lambda \alpha \mu \beta}$. The spacetime admits a group of isometries of dimension four, three, two, one, or no symmetries if, and only if, the connection tensor $H$ obtained in proposition {\em \ref{propo-uE}} fulfills, respectively, the conditions of theorems {\em \ref{theo-G4}, \ref{theo-G3}, \ref{theo-G2}, \ref{theo-G1} or \ref{theo-G0}}. 
\end{corollary}
%


\section{Implementing the algorithm on {\em xAct}. Two examples}
\label{sec-xAct}

The IDEAL quality of the results enables us to implement our characterization algorithm in symbolic programs. Carrying out on {\em xAct} \cite{xAct} the algorithm given in figure \ref{figure-1} and applying it to several metrics is a work in progress that will be presented elsewhere. Here, we only consider two examples that show the usefulness of our approach in two different situations: the Petrov homogeneous vacuum solution, where the $R$-frame is obtained from the algebraic structure of the Weyl tensor and corollary \ref{coro-IDEAL-Weyl} applies, and the Wils conformally flat pure radiation solution, where we need to consider Riemann derivatives to obtain the $R$-frame.


\subsection{The Petrov homogeneous vacuum solution}
\label{subsec-Petrov}

The Petrov homogeneous vacuum solution \cite{Petrov-sol} \cite{kramer} can be characterized as the unique type I vacuum solution with constant Weyl eigenvalues \cite{FS-typeI-a}. It admits a group of isometries G$_4$, and only one Killing vector has its Killing two-form aligned with a Weyl principal bivector \cite{FS-typeI-b}.  

This metric admits a $R$-frame depending algebraically of the Weyl tensor. Thus, we can apply proposition \ref{propo-typeI} to determine the connection tensor, and we can test if it fulfills the conditions of theorem \ref{theo-G4} on the existence of a G$_4$. 

In \cite{notebooks}, you can find a notebook in which we implement all of this. First, we determine the Petrov-Bel type of the metric by following an adapted version of the algorithm proposed in \cite{FMS-Weyl} and we see that, indeed, it is of type I. Then, we obtain its connection tensor from the Weyl tensor as explained in proposition \ref{propo-typeI} and, finally, we follow the steps of the algorithm given in figure \ref{figure-1} to check that it admits a group of isometries G$_4$, namely, that $C^{[1]} = 0$.


\subsection{The Wils conformally flat pure radiation solution}
\label{subsec-Noinvariants}

The conformally flat pure radiation solution  \cite{Wils} is a metric with no symmetries or polynomial invariants \cite{Koutras-Mc}. The metric line element is given by:
\be \label{Wils-metric}
\begin{array}{c}
\dif s^2\! =\! - 2 x\, \dif u  \dif w + 2 w\, \dif u  \dif x - F \dif u^2 + \dif x^2 + \dif y^2, \\[2mm]
 F \equiv \frac12 \phi^2(x^2+y^2)- w^2 , \qquad \phi \equiv 2 \sqrt{|x f(u)|} \, ,
 \end{array}
\ee
where $f(u)$ is an arbitrary real function. Let us consider the null frame $\{ \ell, k, e_2, e_3\}$, 
\be \label{Wils-frame}
\hspace{-12mm} \ell = \frac{1}{2 \phi}[F \dif u + 2 x \dif  w - 2 w \dif x] , \qquad k = \phi \dif u , \qquad e_2 = \dif x , \qquad e_3 = \dif y .
\ee
The Weyl tensor vanishes, and the study of the Ricci tensor $\tilde{R}$ and its derivatives leads to the following relations:
\be \label{Wils-relations}
\begin{array}{c}
\hspace{-12mm} \tilde{R} = k\! \otimes \! k , \qquad x^2 Q = \tilde{R}\! \otimes \! \tilde{R}, \qquad  S^2 = e_3\! \otimes\! e_3, \qquad 2 k \otimes \ell = U - U^2 . \\[2mm]
\hspace{-12mm} Q_{\alpha \beta \lambda \mu} \equiv \nabla_{\alpha} \tilde{R}_{\beta \rho}  \nabla_\lambda \tilde{R}_{\mu}^{\ \rho} , \qquad  S \equiv \frac12 \nabla \dif x^2\! -\! g , \qquad U \equiv *(e_2 \wedge e_3) .
\end{array}
 \ee
From these expressions it follows that the tetrad (\ref{Wils-frame}) defines a $R$-frame. Then, we can obtain the connection tensor $H$ given in (\ref{H-wedge}), with $\sqrt{2} e_0 = \ell + k$ and $\sqrt{2} e_1 = \ell - k$.

Again, in \cite{notebooks} we uploaded a notebook in which we check that this metric is of Petrov-Bel type O, we get its connection tensor from the $R$-frame obtained above and check that it has no symmetries by following the steps of the algorithm in figure \ref{figure-1}. In this case we obtain that $C^{[1]} \neq 0$, $C^{[11]} \neq 0$, $C^{[111]} \neq 0$, $C^{[1111]} = 0$ and $C^{[1112]} \neq 0$.


\section{Discussion and work in progress}
\label{sec-discussion}

In this paper, we have presented an algorithmic characterization of the spacetimes admitting a group G$_r$ of isometries when a $R$-frame is known by using the connection tensor. Moreover, in spacetimes of Petrov-Bel type I, II or III, we have offered an explicit expression of the connection tensor in terms of the Weyl tensor, thus obtaining a fully IDEAL labelling of these geometries.

In spacetimes of Petrov-Bel type O, N or D, obtaining the frame could also require using derivatives of the curvature tensor. In these cases further work is needed to analyze if proposition \ref{propo-uE}, or other criteria that could be stated, apply. 

We have already commented in the previous section that we want to fully implement the algorithm on the {\em xAct Mathematica} suite of packages. But our results induce other new issues to analyze, which we briefly comment below.


\subsection{Derivation order in the characterization conditions. Induced classification}
\label{subsec-derivativeorder}

Knowing the maximum order of derivation necessary to close the Cartan-Karlhede approach is an important issue in the metric equivalence problem. Here, we are not concerned with this question but with obtaining an IDEAL characterization that allows a fully algorithmic analysis. However, it is suitable to know what is the order of derivation of our conditions.

The connection tensor depends on the first derivatives of the Riemann-frame, and concomitants $C^{[q]}(H)$ depend on derivatives of order $q$ of the connection tensor. Consequently, $C^{[q]}(H)$ are $(q\!+\!1)$-differential concomitants of the $R$-frame.  

Then, we have that the whole characterization presented in section \ref{sec-dimension} involves, at most, derivatives of order $5=4+1$ on the $R$-frame: condition G$_4$, second-order derivatives; conditions G$_3$, G$_{2a}$ and G$_{2a}$, third-order derivatives; conditions G$_{2b}$, G$_{1b}$ and G$_{1c}$, fourth-order derivatives; and condition G$_{1d}$, fifth-order derivatives. 

What is the order of derivation with respect to the Riemann tensor? The answer depends on how the $R$-frame has been obtained from the Riemann tensor. When this $R$-frame is determined by the algebraic structure of the Weyl and Ricci tensor, the derivation orders in the $R$-frame and in the Riemann tensor coincide. This occurs, for example, in most of the cases considered in section \ref{sec-computing-H}, and in particular in the vacuum Petrov solution studied in subsection \ref{subsec-Petrov}. 

However, if we need the covariant derivatives of order $p$ on the Riemann tensor to determine the $R$-frame, the derivation order in the Riemann tensor is $p+q$, where $q$ is the order of derivation on the $R$-frame. This is the case of the radiation solution considered in subsection \ref{subsec-Noinvariants}. 

Note that our study induces a classification of the spacetimes admitting a G$_2$ (classes G$_{2a}$ and G$_{2b}$) and a G$_{1}$ (classes G$_{1a}$, G$_{1b}$, G$_{1c}$ and G$_{1d}$). But also induce a classification in eight classes of the spacetimes with no symmetries (corresponding to the eight vertical end arrows in figure \ref{figure-1}).


\subsection{Causal character of the orbits of the isometry group}
\label{subsec-causalOrbites}

Once we know the dimension of the isometry group, we can determine the orbits and their causal character (space-like, time-like or null). The orbits are the $r$-surface orthogonal to the gradient of the $4-r$ independent invariant scalars, and these gradients appear in the first component of the concomitants $C^{[q]}$ involved in the characterization of the G$_r$. In fact, every Killing vector $\xi$ fulfills $i(\xi) C^{[q]} = 0$. 

As an example to show that the $C^{[q]}$ concomitants determine the orbits and their causal character, we analyze the case of a G$_3$. Now, conditions in theorem \ref{theo-G3} hold and $C^{[1]}$ generates only one independent invariant scalar. Then, for any vector $v$ and any 2-form $V$, $w_\lambda = C^{[1]}_{\lambda \rho \mu \nu}v^\rho V^{\mu \nu}$ defines a direction which is collinear with the gradient of this scalar. Consequently, we obtain:
\begin{proposition} \label{signograd}
If a spacetime fulfills conditions of theorem {\em \ref{theo-G3}}, then it admits a transitive  group of isometries G$_3$, and the orbits are the hypersurfaces orthogonal to the vector $w_\lambda = C^{[1]}_{\lambda \rho \mu \nu}v^\rho V^{\mu \nu}$, where $v$ is a vector and $V$ a two-form such that $w \not=0$. The orbits are time-like, space-like or null depending on whether $w^2>0$, $w^2 <0$ or $w^2=0$, respectively.
\end{proposition}
%


\subsection{Characterization of the Abelian {\em G}$_2$ of isometries}
\label{subsec-G2abelia}

The analysis and characterization of the algebra structures for groups G$_2$, G$_3$ or G$_4$ is an issue that needs further work from our approach. For a two-dimensional group, this analysis means that we have to distinguish if the group is Abelian.

When the spacetime admits a G$_2$, the orbits are orthogonal to the gradients of the two independent scalars. If $x, y$ are such scalars, the volume element of these orbits is proportional to the two-form $*(\dif x \wedge \dif y)$. Note that
these invariant scalars appear in two different ways depending on the type G$_{2a}$ or G$_{2b}$ of the group. In the first case, we have $\eta(C^{[1]}, C^{[1]}) \neq 0$ and in the second case we have $\eta(C^{[1]}, C^{[1]}) = 0$ and $\eta(C^{[1]}, C^{[2]}) \neq 0$.

If  $\xi_1$, $\xi_2$ are two Killing fields, from (\ref{nabla-killing}) we obtain:
\be
[\xi_1, \xi_2]_\lambda  = \xi_2^{\alpha} \xi_1^{\beta} (H_{\alpha \beta \lambda}- H_{\beta \alpha \lambda}) \, .
\ee
Then, using the fact that $\eta(C^{[1]}, C^{[1]}) $ or $\eta(C^{[1]}, C^{[2]})  $ are the tensors orthogonal to the invariant scalars, we get:
\begin{proposition}
If a spacetime admits a maximal $G_{2}$ group of isometries with $\eta(C^{[1]}, C^{[1]}) \neq 0$ (respectively, $\eta(C^{[1]}, C^{[1]}) = 0$), then it is commutative if, and only if, $\eta^{\alpha \beta \mu \nu} C^{[1]}_{\alpha \bar{p}}\, C^{[1]}_{\beta \bar{q}} H_{\mu \nu \lambda} =0$ (respectively, $\eta^{\alpha \beta \mu \nu} C^{[1]}_{\alpha \bar{p}}\, C^{[2]}_{\beta \bar{q}} H_{\mu \nu \lambda} =0$).
\end{proposition}
%


\subsection{Characterization of the {\em G}$_3$ Bianchi types, and the {\em G}$_4$ algebras}
\label{subsec-G3G4}

If the group of isometries is three-dimensional, then we have the transitive action of a Bianchi type on the orbits. When these orbits are space-like or time-like, we can detect the Bianchi type by applying our intrinsic approach to the homogeneous three-dimensional Riemannian \cite{FS-K3} or Lorentzian \cite{FS-L3} metrics. In order to do this, we need to consider the induced metric $\gamma = g - (w,w)^{-1} w \otimes w$ ($w$ is that defined in proposition \ref{signograd}), which is Riemaniann or Lorentzian depending on the sign of $(w,w)\not=0$. The IDEAL conditions so obtained must be added to those of theorem \ref{theo-G3}. When the orbits are null, the study of the transitive action of the Bianchi types is a further outstanding task.

Several authors have studied the invariant classification of the four-dimensional algebras \cite{Petrov-llibre, MacCallum-G4}. However, the explicit characterization (in terms of the curvature tensor) of the homogeneous spacetimes where they act is still an open problem. The IDEAL conditions obtained from this study should be added to those of theorem \ref{theo-G4}. 


\subsection{Generalitzation to a $n$-dimensional Riemannian space}
\label{subsec-generalitzacio}

The results that we have presented here for a four-dimensional spacetime can be easily generalized to an $n$-dimensional Riemannian space of any signature. For $n=5$, the characterization theorems for the groups G$_5$, G$_4$, G$_3$ and G$_2$ will be similar to those of theorems \ref{theo-G4}, \ref{theo-G3}, \ref{theo-G2} and \ref{theo-G1} characterizing, respectively, the groups G$_4$, G$_3$, G$_2$ and G$_1$ for $n=4$. And for characterizing a G$_1$ we obtain eight subclasses G$_{1x}$, and in one of them we also need to use the concomitant $C^{[5]}$.

Generically, for an $n$-dimensional Riemannian space, we will have no subclasses for a G$_n$, and for a G$_{n-k}$ we will have $2^{k-1}$, $k= 1...n$, subclasses.

\ack 
This work has been supported by the Spanish Ministerio de Ciencia, Innovaci\'on y Universidades, Project PID2019-109753GB-C21/AEI/10.13039/501100011033, and the Generalitat Valenciana Project AICO/2020/125. S.M. acknowledges financial support from the Generalitat Valenciana (grant CIACIF/2021/028). 


\appendix

\section{The connection tensor in terms of a bivector frame}
\label{sec-bivectors}

A one-to-one correspondence exists between an oriented orthonormal frame of vectors $\{e_0, e_A \}$, $A=1,2,3$, and the orthonormal frame $\{ {\cal U}_A \}$ of the three-dimensional space of bivectors, defined by
\begin{equation} \label{sd-frame}
{\cal U}_A = \frac{1}{\sqrt{2}} (U_A - \ci *U_A ) , \qquad U_A =e_0 \wedge e_A \, ,
\end{equation}
where $*$ denotes  the Hodge dual operator. If we denote by $\cdot$
the contraction of adjacent indexes in the tensorial product and
$\varepsilon_{ABC}$ the Levi-Civita symbol, these bivectors satisfy
\begin{equation} \label{sd-orth}
{\cal U}_A \cdot {\cal U}_A = \frac{1}{2} g , \qquad {\cal U}_A \cdot
{\cal U}_B = - \frac{\ci }{\sqrt{2}} \varepsilon_{ABC}\, {\cal U}_C
, \quad A\neq B \, ,
\end{equation}
where the second equation above induces the inherit orientation in
the space of bivectors.

The connection coefficients $\gamma_{\alpha \beta}^{\mu}$ of the frame $\{e_0, e_A \}$ can be
collected into three {\it complex connection {\rm1}-forms} $\Gamma_A^B
=(\gamma_{AD}^B - \ci \, \varepsilon_{ABC} \gamma_{0D}^C) \theta^D$
($A,B...=1,2,3$) that satisfy  $\Gamma_A^B = - \Gamma_B^A$ and 
\be\nabla {\cal U}_A = \Gamma_A^B \otimes  {\cal U}_B \, . \label{nablaU}
\ee
Then, a straightforward calculation shows that the connection tensor $H$ associated with the frame $\{e_0, e_A \}$ can be obtained in terms of the frame $\{ {\cal U}_A \}$ as
\begin{equation} \label{calH}
\hspace{-10mm} H =\frac{1}{\sqrt{2}}({\cal H} + \bar{\cal H}) \, ,  \qquad {\cal H} =-\frac{1}{\sqrt{2}} \sum_{A=1}^3 \nabla {\cal U}_A \cdot {\cal U}_A = - \ci \! \sum_{(ABC)} \Gamma_A^B \otimes {\cal U}_C \, ,
\end{equation}
where $(ABC)$ denotes cyclic permutation of $(1\,2\,3)$.

A null frame $\{ {\cal U}, \, {\cal L}_+,  \, {\cal L}_-\}$, ${\cal L}_{\pm} = \frac{1}{\sqrt{2}} ({\cal U}_2 \pm \ci \, {\cal U}_3 )$, can also be considered in the space of bivectors. This frame satisfies
\begin{equation} \label{ornull}
{\cal L}_{\pm} \cdot {\cal U}=  \pm \frac{1}{\sqrt{2}} {\cal L}_{\pm}, \qquad  {\cal L }_{\pm} \cdot
{\cal L}_{\mp}= \frac{1}{2} g \mp \frac{1}{\sqrt{2}}\, {\cal U} \, .
\end{equation}
In terms of such a null frame, the self-dual connection tensor ${\cal H}$ can be obtained as
\begin{equation} \label{SDnullct}
{\cal H} = - \frac{1}{\sqrt{2}} \left(  \nabla {\cal U} \cdot {\cal U} +  \nabla {\cal L}_+\! \cdot \! {\cal L}_- +  \nabla {\cal L}_- \! \cdot \! {\cal L}_+  \right) \, .
\end{equation}
%



\section*{References}
\begin{thebibliography}{999}


\bibitem{eisenhart-33}  Eisenhart L P 1933 {\em Continuous groups of transformations} (Princeton University Press, Princeton)

\bibitem{kramer} Stephani H, Kramer D, MacCallum M A H, Hoenselaers C and Herlt E 2003 {\it Exact Solutions of the Einstein`s Field Equation} (Cambridge University Press, Cambridge)

\bibitem{cartan} Cartan E 1946 {\it Le\c{c}on sur le G\'eometrie des Espaces de Riemann} 2nd edn (Paris: Gauthier-Villars)

\bibitem{brans} Brans C H 1965 {\it J. Math. Phys.} {\bf 6} 94

\bibitem{karlhede} Karlhede A 1980 {\it Gen. Relativ. Gravit.} {\bf 12} 693


\bibitem{karlhede-maccallum} Karlhede A and MacCallum M A H 1982 {\it Gen.
Relativ. Gravit.} {\bf 14} 673

\bibitem{MacCallum-2015} MacCallum M A H 2015 Spacetime invariants and their uses {\it Proc. Int. Conf. Relativ. Astrophys.} 122

\bibitem{tomoda} Nozawa M and Tomoda K, 2019 {\it Class. Quantum
Grav.} {\bf 36} 155005

\bibitem{FS-K3} Ferrando J J and S\'aez J A 2021 {\it Class. Quantum Grav.}
{\bf 38} 067001

\bibitem{FS-G3} Ferrando J J and S\'aez J A 2020 {\it Class. Quantum Grav.}
{\bf 37} 185011

\bibitem{FS-L3} Ferrando J J and S\'aez J A 2022 {\it Class. Quantum Grav.}
{\bf 39} 165014

\bibitem{Kerr-62} Kerr R P 1962 {\it Tensor} {\bf 12} 74 

\bibitem{Kerr-63} Kerr R P 1963 {\it Phys. Rev. Lett.} {\bf 11} 237 

\bibitem{FMS-Weyl} Ferrando J J, Morales J A and S\'aez J A 2001 {\it Class. Quantum Grav.}
{\bf 18} 4939

\bibitem{Petrov-sol} Petrov A Z 1962 Gravitational field geometry as the geometry of automorphisms {\em Recent Developments in General Relativity} (Oxford: Pergamon) PWN 379 

\bibitem{Wils} Wils P 1989 {\it Class. Quantum Grav.} {\bf 6} 1243

\bibitem{Defrise} Defrise L 1969 {\it Groupes d'isotropie et groupes de stabilit\'e conforme dans les espaces
lorentziens} Thesis, Universit\'e Libre de Bruxelles 

\bibitem{FS-SHC} Ferrando J J and S\'aez J A 2023 {\it Spatially-Homogeneous Cosmologies} in preparation

\bibitem{xAct} Mart\'in-Garc\'ia J M \textit{xAct: Efficient tensor computer algebra for the Wolfram Language} http://www.xact.es/

\bibitem{FS-typeI-a} Ferrando J J and S\'aez J A 2003 {\it Class. Quantum Grav.} {\bf 20} 5291 

\bibitem{FS-typeI-b} Ferrando J J and S\'aez J A 2006 {\it J. Math. Phys.} {\bf 20} 112501 

\bibitem{notebooks} Mengual S https://github.com/SalvaMS6/Dimension-of-the-isometry-group


\bibitem{Koutras-Mc} Koutras A and McIntosh C 1996 {\it Class. Quantum Grav.} {\bf 13} L47

\bibitem{Petrov-llibre} Petrov A Z 1969 {\em Einstein Spaces} (Pergamon Press, Oxford) 

\bibitem{MacCallum-G4} MacCallum M A H 1999 {\it On the classification of the real four-dimensional Lie algebras} in On Einstein's path: Essays in honor of Engelbert Schucking, ed. A L Harvey (springer Velag, New York)


\end{thebibliography}
\end{document}